\documentclass[aps,prb,twocolumn,groupedaddress,showpacs]{revtex4}

\usepackage{graphicx}

\begin{document}

\title{Photoemission study of the metal-insulator transition in
VO$_2$/TiO$_2$(001) : Evidence for strong
electron-electron and electron-phonon interaction}

\author{K.~Okazaki$^{1,}$\cite{adr}, H.~Wadati$^1$, A.~Fujimori$^{1,2}$,
M.~Onoda$^3$, Y.~Muraoka$^4$ and Z.~Hiroi$^4$}

\affiliation{$^1$Department of Physics, University of Tokyo, Bunkyo-ku,
Tokyo 113-0033, Japan} 
\affiliation{$^2$Department of Complexity Science and Engineering,
University of Tokyo, Bunkyo-ku, Tokyo 113-0033, Japan}
\affiliation{$^3$Institute of Physics, University of Tsukuba, Tsukuba,
Ibaraki 305-8571, Japan} 
\affiliation{$^4$Institute for Solid State Physics, University of Tokyo,
Kashiwa, Chiba 277-8581, Japan}
 
\date{\today}

\begin{abstract}
 We have made a detailed temperature-dependent photoemission study of
 VO$_2$/TiO$_2$(001) thin films, which show a metal-insulator transition
 at $\sim$ 300 K. Clean surfaces were obtained by annealing the films in
 an oxygen atmosphere. Spectral weight transfer between the coherent and
 incoherent parts accompanying the metal-insulator transition was
 clearly observed. We also observed a hysteretic behavior of the spectra
 for heating-cooling cycles. We have derived the ``bulk'' spectrum of
 the metallic phase and found that it has a strong incoherent part. The
 width of the coherent part is comparable to that given by
 band-structure calculation in spite of its reduced spectral weight,
 indicating that the momentum dependence of the self-energy is
 significant. This is attributed to by ferromagnetic fluctuation
 arising from Hund's rule coupling between different $d$ orbitals as
 originally proposed by Zylbersztejn and Mott. In the insulating phase,
 the width of the V $3d$ band shows strong temperature dependence. We
 attribute this to electron-phonon interaction and have reproduced it
 using the independent boson model with a very large coupling constant.
\end{abstract}

% insert suggested PACS numbers in braces on next line
% insert suggested keywords - APS authors don't need to do this
%\keywords{}

%\maketitle must follow title, authors, abstract, \pacs, and \keywords
\maketitle

\section{Introduction}\label{Intro} 

To elucidate the physical properties of Mott-Hubbard systems has been
one of the most challenging subjects in condensed matter physics and has
continued to pose controversial theoretical as well as experimental
issues. In this respect, the single-particle spectral function of
transition-metal oxides, particularly of early transition-metal oxides,
is of fundamental importance in the physics of Mott-Hubbard systems. In
fact, the coexistence of the coherent part [the quasi-particle (QP)
band] and the incoherent part (a remnant of the upper and lower Hubbard
bands) in the spectral function and spectral weight transfer between
them as a function of electron correlation strength $U/W$, where $U$ is
the Coulomb repulsion energy and $W$ is the bandwidth, are a remarkable
manifestation of electron correlation as identified in the photoemission
spectra of V and Ti oxides.~\cite{Fujimori,Inoue} Theoretically,
dynamical mean-field theory (DMFT) applied to the Hubbard model has
successfully reproduced these characteristic features.~\cite{DMFT}
Comparison between experiment and theory, however, has not been
straightforward. In DMFT, the self-energy is necessarily local, and
therefore the density of states (DOS) at the Fermi level ($E_{\rm F}$),
namely, the spectral intensity at $E_{\rm F}$ remains the same as that
of the non-interacting system and electron correlation effect manifests
itself as the narrowing of the coherent QP band. Experimentally, the
reduction of the DOS at $E_{\rm F}$ rather than the narrowing of the
coherent part was observed in the photoemission spectra of
Ca$_{1-x}$Sr$_x$VO$_3$, suggesting the importance of the momentum
dependence of the self-energy.~\cite{Fujimori,Inoue} However, a recent
``bulk-sensitive'' photoemission study of the same compounds has shown
that the QP band is indeed narrowed and the DOS at $E_{\rm F}$ remains
unchanged from that predicted by band-structure calculation, indicating
that the self-energy is nearly momentum-independent as assumed in
DMFT.~\cite{Sekiyama} In the present work, we have studied another
typical Mott-Hubbard system VO$_2$, which undergoes a metal-insulator
transition as a function of temperature, and examined the spectral
function of the ``bulk'' to see whether the self-energy is
momentum-independent or not. Our results have shown that the self-energy
is {\it momentum-dependent}, probably due to ferromagnetic fluctuations
arising from the multi-orbital nature of the V $3d$ band of the rutile
structure.

VO$_2$ is well-known for its first-order metal-insulator transition
(MIT) at 340 K.~\cite{Morin} The transition is accompanied by a
structural transition. In the high temperature metallic phase it has a
rutile structure while in the low temperature insulating phase ($M_1$
phase) the V atoms dimerize along the $c$-axis and the dimers twist,
resulting in a monoclinic structure. The magnetic susceptibility changes
from paramagnetic to nonmagnetic in going from the metallic to the
insulating phases. Hence, this transition is analogous to a Peierls
transition and in fact the importance of electron-phonon interaction has
been demonstrated by Raman scattering~\cite{Raman} and x-ray
diffraction~\cite{XRD} studies. On the basis of local-density
approximation (LDA) band-structure calculation, Wentzcovitch {\it et
al.}~\cite{Wentzcovitch} concluded that the insulating phase of VO$_2$
is an ordinary band (Peierls) insulator. On the other hand, the magnetic
susceptibility of the high-temperature metallic phase is unusually high
and temperature dependent, indicating the importance of
electron-electron correlation. Furthermore, Cr-doped VO$_2$ or pure
VO$_2$ under uniaxial pressure in the [110] direction of the rutile
structure has another monoclinic insulating phase called $M_2$ phase. In
the $M_2$ phase, half of the V atoms form pairs and the other half form
zig-zag chains.~\cite{Marezio} While the V atoms in the pairs are
nonmagnetic, those in the zig-zag chains have local moment and are
regarded as one-dimensional Heisenberg chains according to an NMR
study~\cite{Pouget}. Based on these observations for the $M_2$ phase,
Rice {\it et al.}~\cite{Rice} objected Wentzcovitch {\it et al.}'s
argument. Thus it still remains highly controversial whether the MIT of
VO$_2$ is driven by electron-phonon interaction (resulting in a Peierls
insulator) or electron-electron interaction (resulting in a Mott
insulator forming a spin-Peierls-like state).

To deal with the above problems, photoemission spectroscopy is a
powerful technique and in fact has been extensively applied to this
material.~\cite{Wertheim, Blaauw, Sawatzky, Shin, Bermudez, Goering,
Okazaki} However, detailed photoemission studies of the MIT has been
hampered because the transition temperature of bulk VO$_2$ is rather
high and therefore it is difficult to keep the surface clean in an
ultra-high vacuum for the high-temperature metallic phase. Also, because
the transition is strongly first-order with the structural change, the
sample is destroyed when it crosses the MIT and one can go through the
transition only once for one sample.

In this work, we have avoided those experimental difficulties by using
thin film samples epitaxially grown on TiO$_2$(001) surfaces using
the pulsed laser deposition technique.~\cite{Muraoka} After having
obtained a clean surface by oxygen annealing, the surface remained
fairly stable for several hours even in the high-temperature metallic
phase and allowed us to study detailed temperature-induced changes both
in the metallic and insulating phases including the hysteretic behavior
across the MIT. Spectral weight transfer between the coherent and
incoherent parts of the V $3d$ spectral function accompanying the
metal-insulator transition was clearly observed. We have attempted to
deduce the ``bulk'' photoemission spectrum by subtracting surface
contributions from the measured spectra. We compare the ``bulk''
spectrum with the band-structure calculation to discuss electron
correlation in the metallic phase, particularly possible momentum
dependence of the self-energy. We have also found a strong temperature
dependence in the spectra of the insulating phase and attributed it to
strong electron-phonon interaction.

\section{Experimental}
VO$_2$/TiO$_2$(001) thin films were prepared using the pulsed laser
deposition technique as described in Ref.~\onlinecite{Muraoka}. A
V$_2$O$_3$ pellet was used as a target. During the deposition, the
substrate temperature was kept at 733 K and oxygen pressure was
maintained at 1.0 Pa. The film thickness was about 10 nm. The epitaxial
growth was confirmed by four-cycle x-ray diffraction (XRD) measurements
and the MIT was confirmed by electrical resistivity measurements showing
a jump of about thee orders of magnitude. The transition temperature in
the films was 295 K on heating cycle and 285 K on cooling cycle, while
the MIT occurs at 340 K in bulk samples.~\cite{Morin} The reduced MIT
temperature of the films is due to the compressive strain from the
TiO$_2$ substrate.~\cite{Muraoka} Bulk VO$_2$ single crystals were
prepared by the chemical vapor transport method and measured as
described in Ref.~\onlinecite{Okazaki}.

X-ray photoemission spectroscopy (XPS) and ultra-violet photoemission
spectroscopy (UPS) measurements were performed using the Mg $K\alpha$
line ($h\nu$ = 1253.6 eV) for XPS and the He I resonance line ($h\nu$ =
21.2 eV) for UPS with a VSW hemispherical analyzer. Estimation of the
instrumental resolution and binding energy calibration were made by
measuring gold spectra. The total energy resolution was $\sim$ 0.8 eV
for XPS and $\sim$ 30 meV for UPS. Clean surfaces were obtained by
annealing the films in a preparation chamber connected to the
spectrometer at 643 K under $\sim$ 1 Pa oxygen atmosphere for 1 hour prior
to the photoemission measurements. 

\section{Results}
\begin{figure}[t]
\begin{center}
\includegraphics[width=9cm]{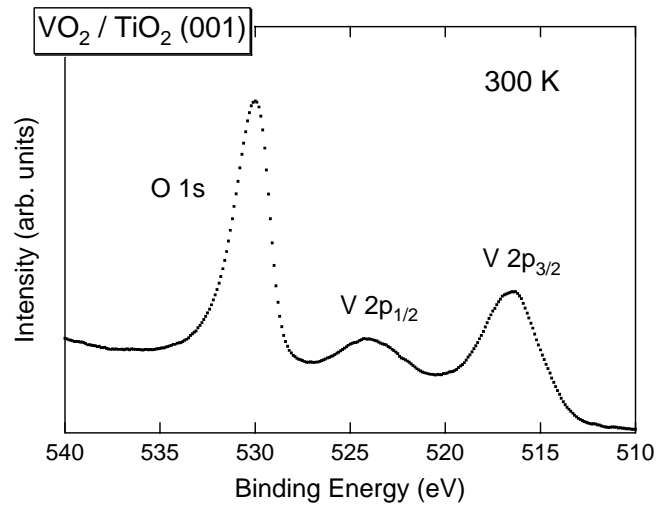}
\caption{\label{Fig1}O $1s$ and V $2p$ core-level XPS spectra of
 VO$_2$/TiO$_2$(001) thin film.}
\end{center}
\end{figure}

Figure~\ref{Fig1} shows the O $1s$ and V $2p$ core-level XPS spectra of
a VO$_2$/TiO$_2$(001) film taken at room temperature. Contributions
from the Mg $K\alpha_3$ and $K\alpha_4$ satellites have been
subtracted. The O 1s peak shows a single peak without any contamination
signals at higher binding energies, indicating that the surface became
sufficiently clean by the above procedure.

\begin{figure}[t]
\begin{center}
\includegraphics[width=8cm]{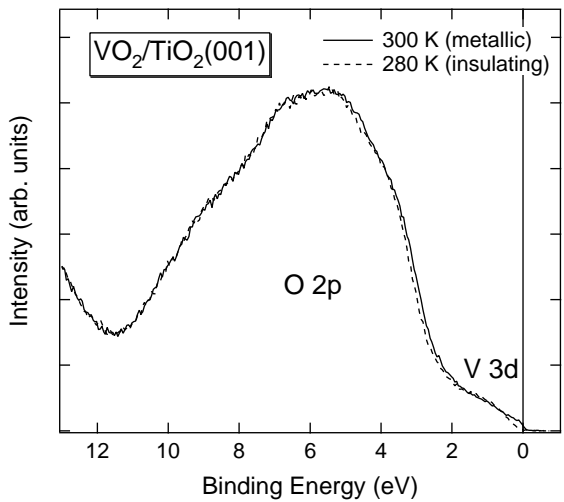} 
\caption{\label{Fig2}Valence-band photoemission spectra of
 VO$_2$/TiO$_2$(001) thin film. Secondary electron background has been
 subtracted.}
\end{center}
\end{figure}
\begin{figure}[t]
\begin{center}
\includegraphics[width=7.8cm]{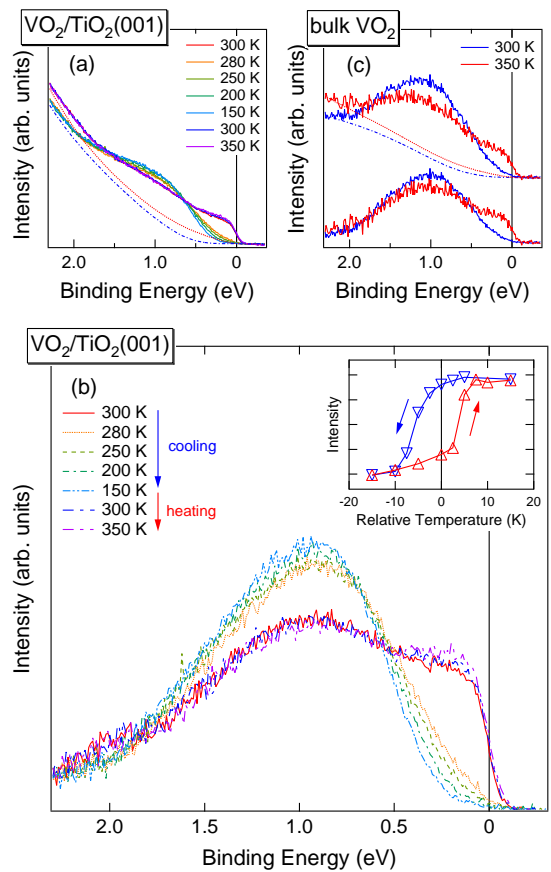} 
\caption{\label{Fig3}Photoemission spectra of VO$_2$/TiO$_2$(001) thin
 film in the V $3d$ band region. (a) Raw data. Dotted and dot-dashed
 lines indicate estimated contributions from the tail the O $2p$
 band. (b) V $3d$ band with the O $2p$ contributions being
 subtracted. The inset shows a hysteretic behavior of the spectral
 intensity around $E_{\rm F}$ across the MIT. (c) Photoemission spectra
 of a bulk single crystal VO$_2$. Top: Raw data; Bottom: Background
 being subtracted.}
\end{center}
\end{figure}

Figure~\ref{Fig2} shows the valence-band UPS spectra of
VO$_2$/TiO$_2$(001) taken at 300 K and 280 K. The secondary electron
background has been subtracted following the procedure of Li and
Henrich.~\cite{Henrich} The structures from binding energies $E_B$
$\simeq$ 2 to 12 eV are due to the O $2p$ band. The region from $E_B$
$\simeq$ 2 eV to $E_{\rm F}$ is the V $3d$ band. While the O $2p$ band
shows no clear temperature dependence, the V $3d$ band shows a
temperature dependence, indicating that the MIT occurred between 280 and
300 K.

Figure~\ref{Fig3} (a) and (b) shows the UPS spectra of
VO$_2$/TiO$_2$(001) in the V $3d$ band region taken at various
temperatures. First the temperature was decreased from 300 K to 150 K,
then it was increased again to above 300 K. The spectra showed good
reproducibility within this temperature cycle. The tail of the O $2p$
band has been subtracted as shown in Fig.~\ref{Fig3} (a) and the
resulting spectra have been normalized to the integrated intensity from
$E_B$ = -- 0.3 to 2.3 eV, as shown in Fig.~\ref{Fig3}(b). The MIT is
clearly seen as the change in the spectral intensity at $E_{\rm F}$
accompanied by the spectral weight transfer between the low
binding-energy region $E_B$ = 0-0.5 eV (coherent part) and the high
binding-energy region $E_B$ = 0.5-2 eV (incoherent part). The spectra of
the films (Fig.~\ref{Fig3}(b)) are almost identical to that of the bulk
sample (Fig.~\ref{Fig3}(c)) except for the somewhat stronger temperature
dependence of the film spectra. Because of the much higher stability of
the film surface in vacuum, we could obtain the spectra with much higher
signal-to-noise ratio at much smaller temperature intervals in the whole
temperature range. Also, the film retained the original spectra after
several temperature cycles whereas the bulk single crystal broke into
pieces once it crossed the MIT. In the previous study of bulk single
crystals,~\cite{Okazaki} detailed temperature-dependent studies were
therefore limited to the low-temperature insulating phase, where the
crystal did not break. The present temperature dependence in the
insulating phase has reproduced the bulk crystal results.

Here, we have observed a hysteretic behavior in the spectra with
temperature across the MIT, as demonstrated by the temperature
dependence of the intensity around $E_{\rm F}$ (integrated from $E_B$ =
-- 0.2 to 0.5 eV) shown in the inset of Fig.~\ref{Fig3} (b). The
hysteretic behavior is seen over the temperature range of $\sim$ 10
K. This may indicate the co-existence of metallic and insulating regions
with changing volume fractions in this temperature range. Photoemission
spectra of another system which shows a first-order temperature-induced
metal-insulator transition, $R$NiO$_3$, also shows a similarly gradual
temperature dependence.~\cite{Ni1,Ni2} In the case of $R$NiO$_3$, the
hysteretic behavior in the transport and thermodynamic properties
extends over a wide temperature range of several tens K,~\cite{Ni3} and
correspondingly the photoemission spectra show a more gradual
temperature dependence over the wider temperature range.

\begin{figure}[t]
\begin{center}
\includegraphics[width=7.8cm]{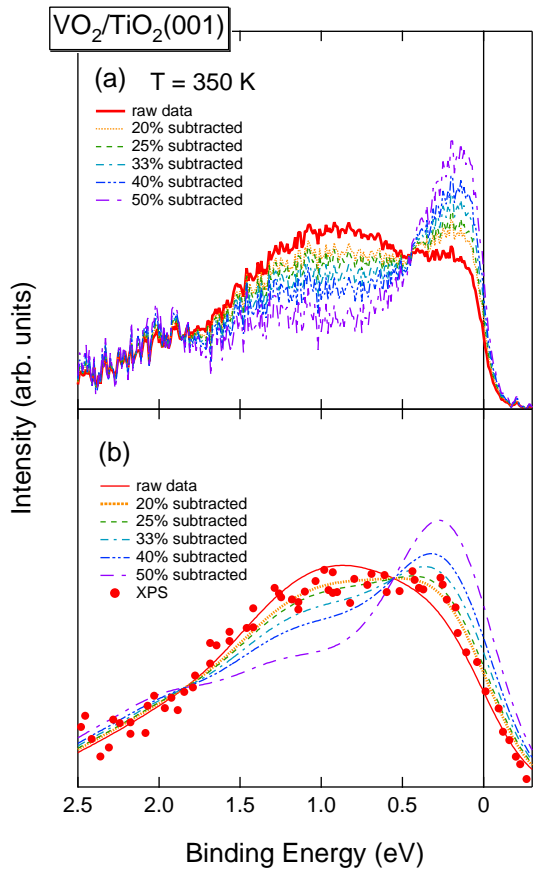} 
\caption{\label{Fig4}(a) Subtraction of ``surface'' contributions with
 various ratios from the photoemission spectrum of the metallic
 phase. (b) Comparison of the spectra in (a) with the XPS spectrum from
 Ref.~\onlinecite{Wertheim}. Here, the curves in (a) have been broadened
 with a Gaussian to account for the lower energy resolution of XPS.}
\end{center}
\end{figure}

Recently, it has been demonstrated that photoemission spectra of
transition-metal oxides are strongly influenced by surface
contributions. In particular, the incoherent part centered around $E_B$
= 1-1.5 eV contains more surface contributions than the coherent
part.~\cite{Maiti, Sekiyama} Since the surface layer tends to have a
smaller bandwidth $W$ and hence a larger $U/W$, the surface layer tends
to have stronger incoherent part and weaker coherent part than in the
bulk.~\cite{Maiti,Liebsch} In order to remove such surface contributions
and to analyze the electronic structure of bulk materials, we attempted
to deduce the ``bulk'' spectrum in the metallic phase under several
assumptions. If the thickness of the surface layer is $d$ and the
photoelectron escape depth is $\lambda$, the observed spectrum
$F(\omega)$ is given by
\begin{equation}
 F(\omega)=F^s(\omega)(1-e^{-d/\lambda})+F^b(\omega)e^{-d/\lambda},
\end{equation}
where $F^s(\omega)$ and $F^b(\omega)$ are the ``surface'' and the
``bulk'' spectra, respectively. As the electronic properties of the
surface layer of VO$_2$ are not precisely known, in the following
analyses, we consider the extreme case where the photoemission spectra
of the surface layer largely consists of the incoherent part. Therefore,
the intensity of the incoherent part in the ``bulk'' spectrum deduced
below should be regarded as the lower bound for the actual bulk
one. 

Assuming that the spectrum at 280 K ($< T_{MI}$) represents the
``surface'' spectrum, because the photoelectron escape depth $\lambda$
is not precisely known, we subtracted the ``surface'' contribution with
various ratios as shown in Fig.~\ref{Fig4} (a). Here, ``20\%
subtracted'' means that the intensity of the subtracted insulator-like
spectrum was 20\% of the total intensity. In Fig.~\ref{Fig4} (b), we
compare the ``bulk sensitive'' XPS spectrum in the
literature~\cite{Wertheim} with those spectra in (a) after having
broadened them with a Gaussian function corresponding to the lower
experimental energy resolution of XPS ($\sim 0.55$ eV). One can see that
the XPS spectrum is in best agreement with ``20\% subtracted''
spectrum. According to the ``universal curve'' of the photoelectron
escape depth,~\cite{Brundle} the escape depth is $\sim$ 10 {\AA} for
$h\nu$ = 21.2 eV and $\sim$ 15 {\AA} for XPS ($h\nu$ = 1486.6 eV),
respectively. Using the surface layer thickness of $2c$ (= 5.7 {\AA}) on
the (001) surface of the rutile structure, where $c$ is the $c$-axis
lattice constant, the surface contribution is $\sim$ 44\% for $h\nu$ =
21.2 eV and $\sim$ 32\% for $h\nu$ = 1486.6 eV, respectively. Hence, the
``20\% subtracted'' spectrum should have 30\% surface contribution and
therefore it is reasonable that the ``20\% subtracted'' spectrum is in
agreement with the XPS spectrum. After all, we consider that the
``bulk'' spectrum is given by the ``40\% subtracted'' spectrum.

\section{Discussion}

\subsection{Metallic phase}

\begin{figure}[t]
\begin{center}
\includegraphics[width=8cm]{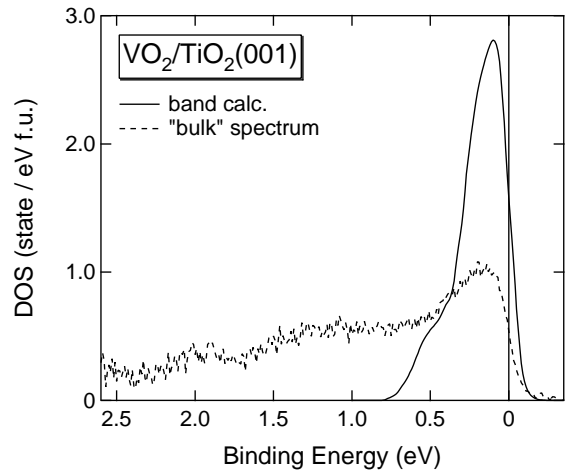} 
\caption{\label{Fig5} ``Bulk'' spectrum is compared with the LDA
 band-structure calculation from Ref.~\onlinecite{Nikolaev}.}  
\end{center}
\end{figure} 

In Fig.~\ref{Fig5}, we compare the ``bulk'' spectrum thus deduced with
the LDA band-structure calculation.~\cite{Nikolaev} The calculated
density of states (DOS) is multiplied by the Fermi-Dirac distribution
function at 350 K and broadened with a Gaussian corresponding to the
instrumental energy resolution. The width of the coherent part $\sim$
0.5 eV is similar to that given by the band-structure calculation. This
means that $m^*/m_b$ is $\sim$ 1, where $m^*$ is the average effective
mass of the QP and $m_b$ is the average bare band mass of the V $3d$
band. From the spectral weight ratio of the coherent part to the
incoherent part, $1:2$, the renormalization factor becomes $Z \sim
1/3$. From the ratio of the intensity at $E_{\rm F}$ of the spectra to
the DOS given by the band-structure calculation, one can estimate the
average $k$-mass defined by $m_k/m_b =
\bigl|\frac{\partial\varepsilon_k} {\partial k}|_{k=k_F}\bigr|/
\bigl|\frac{\partial\varepsilon_k}{\partial k}|_{k=k_F}+
\frac{\partial{\rm Re} \Sigma(k,\omega)}{\partial k}|_{k=k_F}\bigr|$, as
$m_k/m_b\sim$ 0.3. Thus, we again obtain $m^*/m_b = (1/Z)(m_k/m_b) \sim$
1. Note that we have assumed that the surface layer is without the
coherent part around $E_{\rm F}$. If the surface spectrum has a finite
coherent part, the coherent part of the ``bulk'' spectrum would be
further reduced and the deduced $k$-mass would be even smaller. Hence,
$m_k/m_b \sim 0.3$ deduced here should be taken as the upper limit for
$m_k/m_b$, and therefore on can safely conclude the $m_k/m_b$ is smaller
than unity. That is, the momentum dependence of the self-energy is not
negligible in the metallic phase of VO$_2$.

Our finding $m_k/m_b < 1$ is contrasted with the ``bulk'' spectra of
Sr$_{1-x}$Ca$_x$VO$_3$ reported by Sekiyama {\it et al}.~\cite{Sekiyama}
In their result, the width of the coherent part of the V $3d$ band of
SrVO$_3$ is reduced by a factor of $\sim$ 40 \% compared with the LDA
result, and the spectral weight of the coherent part is also reduced by
$\sim$ 40\% due to the transfer of spectral weight to the incoherent
part, resulting in almost the same spectral intensity at $E_{\rm F}$ as
that of the LDA value, that is, the self-energy has negligible momentum
dependence in Sr$_{1-x}$Ca$_x$VO$_3$. The origin of the momentum
dependence of the self-energy in VO$_2$ is not known at this moment. One
possibility is poor screening of long-range Coulomb interaction, but it is
unlikely that the screening effect is so different between VO$_2$ and
Sr$_{1-x}$Ca$_x$VO$_3$. Another possible origin of the momentum
dependence is ferromagnetic fluctuations in VO$_2$ considering the
enhanced magnetic susceptibility in the metallic phase, corresponding to
$m^*/m_b \sim 6$ (Ref.~\onlinecite{susceptibility}). Recently, Liebsch
and Ishida~\cite{Liebsch2} proposed that the multi-orbital nature of the
V $3d$ band is important to describe the metallic phase of VO$_2$
(Ref.~\onlinecite{Laad, Liebsch2}), following the idea originally
proposed by Zylbersztejn and Mott.~\cite{Zylbersztejn} According to
these works, the occupancy of the $t_{2g}$ orbitals is very different
between the metallic and the insulating phases due to lattice distortion
in the insulating phase and this affects the role of local Coulomb
interaction in each phase. If many orbitals contribute to the metallic
conductivity, ferromagnetic correlation may arise from Hund's coupling
between those orbitals. Further studies are necessary to confirm this
scenario.

\subsection{Insulating phase}

As stated in Sec.~\ref{Intro}, it has been controversial whether the
insulating band gap is primarily caused by the lattice distortion or the
electron-electron interaction. In order address this issue, it is useful
to examine the photoemission spectral line shape of the V $3d$ band in
the insulating phase.

\begin{figure}[t]
\begin{center}
\includegraphics[width=8cm]{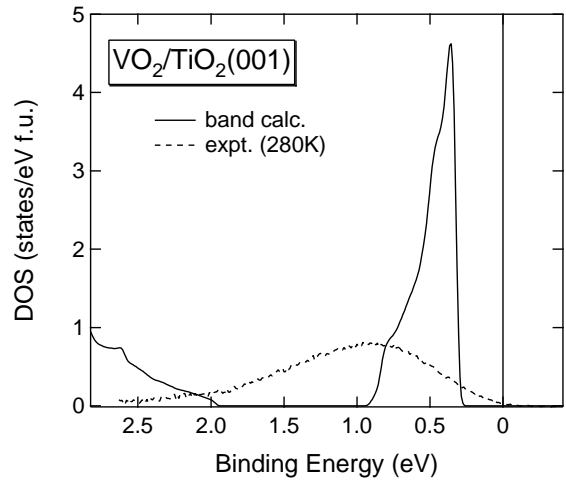}
\caption{\label{Fig6} UPS spectrum in the insulating phase (280 K)
 compared with the LDA$+U$ band-structure calculation ($U = 4.0$
 eV).~\cite{LDA+U}}
\end{center}
\end{figure}

In Fig.~\ref{Fig6}, we compare the UPS spectrum of the insulating phase
with the DOS given by the LDA$+U$ band-structure calculation ($U$ = 4.0
eV) by Huang {\it et al.},~\cite{LDA+U} where the effect of
electron-electron interaction is taken into account on the mean-field
(i.e. Hartree-Fock) level. The figure shows that, although the
insulating gap can be produced by the LDA$+U$ method, the Gaussian-like
broad spectrum of V $3d$ band cannot be reproduced. This disagreement
between theory and experiment cannot be explained by a surface effect
because since the bulk component should have finite contribution ($\sim$
60 \%) even for this low photon energy and therefore the sharp feature
should be clearly visible overlapping the surface signals. This means
that electron-electron interaction on the mean-field level is not
sufficient to understand the photoemission spectrum of the insulating
VO$_2$. Here, we point out that the photoemission spectra of a
double-layer manganite La$_{1.2}$Sr$_{1.8}$Mn$_2$O$_7$ also show a broad
Gaussian-like line shape in striking contrast with LDA band-structure
calculation, and this discrepancy has been attributed to the strong
electron-phonon coupling.~\cite{Mn} Furthermore, one can recognize that
the spectra of VO$_2$ show very strong temperature dependence in the
insulating phase as shown in Fig~\ref{Fig3}. The same temperature
dependence was observed in the spectra of bulk single crystal and was
attributed to electron-phonon interaction, because the core-level
spectra also show a similarly strong temperature
dependence.~\cite{Okazaki}

\begin{figure}[t]
\begin{center}
\leavevmode 
\includegraphics[width=7.5cm]{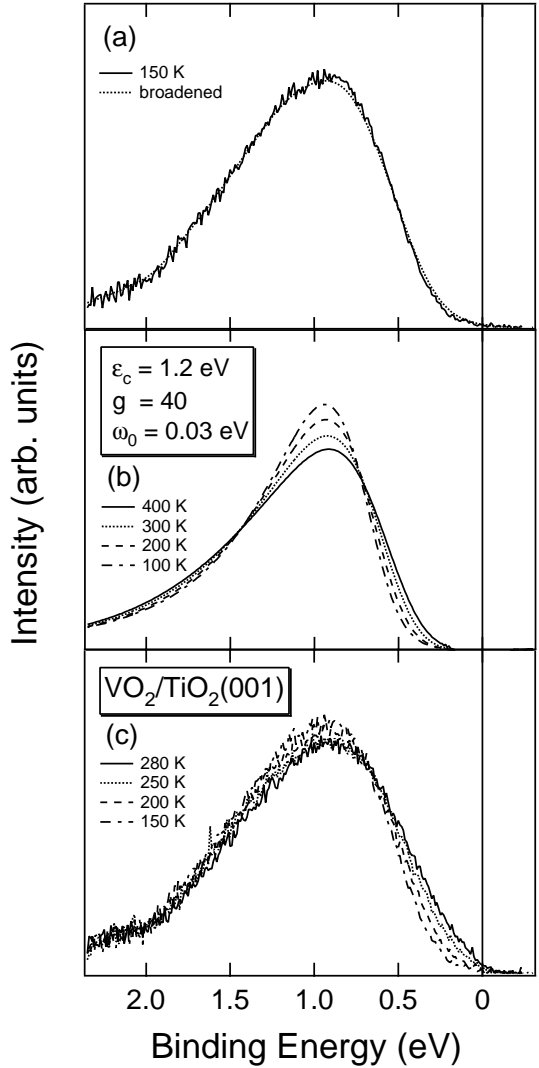}
\caption{\label{Fig7}Comparison of the spectral function of the
independent boson model with the photoemission spectra of
VO$_2$/TiO$_2$(001) in the insulating phase. (a) Spectrum at 150 K and
its Gaussian-broadened (FWHM $=$ 200 meV) spectrum, (b) Independent
boson model calculation, (c) Measured photoemission spectra [the same as
Fig.~\ref{Fig3}].}
\end{center}
\end{figure}

Simple thermal broadening cannot explain the observed temperature
dependence as shown in Fig.~\ref{Fig7} (a), where the photoemission
spectrum taken at 150 K and the same spectrum broadened with a Gaussian,
whose FWHM is 200 meV ($\sim$ $8k_{\rm B}T$ at 300 K) are compared. In
order to simulate the temperature-dependence based on the
electron-phonon interaction mechanism, we introduce the spectral
function of the independent boson model at finite
temperature given by~\cite{Mahan}
\begin{eqnarray*} 
 A(\omega)=e^{-g(2N+1)}\sum_l\frac{g^l}{l!}\sum_{m=0}^l{}_l C_mN^m(N+1)
^{l-m} \\
\times \delta(\omega-\varepsilon_c+\Delta-(l-2m)\omega_0),
\end{eqnarray*}
where $N$ is the phonon occupation number, $g$ is the electron-phonon
coupling constant, $\omega_0$ is the average phonon energy, the $\Delta
= g\omega_0$ is the electron self-energy shift due to coupling to
phonons and $\varepsilon_c$ gives the peak position of the spectra. This
spectral function describes the situation where spectral weight is
transfered from the zero-phonon line at the lowest binding energy
$\varepsilon_c-\Delta$ to $\varepsilon_c-\Delta+(l-2m)\omega_0$ by
emitting $(l-m)$ phonons or absorbing $m$ phonons.  Because this
spectral function consists of a series of delta functions, one has to
broaden this function to compare with the photoemission spectra. For
this purpose, we use a simplified model self-energy
$\Sigma(\omega)=G\omega/(\omega+i\gamma)^2$, which simulates the ``life
time'' broadening which increases with increasing binding
energy.~\cite{Saitoh} In Fig.~\ref{Fig7} (b) and (c), we compare the
spectral function of the independent boson model with the photoemission
spectra of VO$_2$/TiO$_2$(001) in the insulating phase. We have chosen
the parameters as $\varepsilon_c=1.2$ eV, $g=40$ and $\omega_0=0.03$
eV (Ref.~\onlinecite{Gervais}) for the independent boson model and $G=1.5$ eV$^2$ and
$\gamma=2$ eV for the broadening. Because the tail of the lower binding
energy side of the photoemission spectrum reaches the vicinity of
$E_{\rm F}$, we have chosen the coupling constant $g$ so that
$\varepsilon_c-\Delta \sim 0$. Hence, $g=\varepsilon_c/\omega_0 \approx
40$. One can see that the temperature-dependent spectral function of the
independent boson model qualitatively reproduces the photoemission
spectra.

The extremely large coupling constant $g=40$ may appear unphysical at
first sight. However, Citrin {\it et al.}~\cite{Citrin} have shown that
in ionic crystals the coupling constant can be large and reported $g =
55$ for KI based on their analysis of temperature-dependent core-level
photoemission spectra. They have estimated the value of $g$ using the
expression $g = e^2(6/\pi V)^{1/3}(1/\varepsilon_\infty
-1/\varepsilon_0) / \omega_0$, where $V$ is the volume of a primitive
unite cell, $\varepsilon_\infty$ and $\varepsilon_0$ are the high- and
low-frequency limits of the dielectric constant. For VO$_2$, $V \sim$
29.5 \AA$^3$, $\varepsilon_\infty$ and $\varepsilon_0$ have been given
as 5 and 30 in Ref.~\onlinecite{Zylbersztejn}, respectively. From these
values, $g$ is calculated as $\sim$ 32. On the other hand, Egdell {\it
et al.}~\cite{Egdell} have used $1/2R$ in stead of $(6/\pi V)^{1/3}$ for
perovskite-type V oxides, where $R$ is the ionic radius of the
transition-metal ion. In this case, $g$ becomes as large as 47. ($R$ is
0.82 {\AA} for VO$_2$.~\cite{Gervais}) Hence, we can say that $g =
30-50$ is a reasonable value for the electron-phonon coupling constant
for VO$_2$.

Finally, we comment on how the strong electron-electron interaction and
the strong electron-phonon coupling are related with each other in
VO$_2$ (and probably in other transition-metal oxides as well). In ionic
crystals, the on-site Coulomb repulsion energy $U$ is given by $ U =
e^2/2 \varepsilon_\infty R$. Therefore, according to Egdell {\it et
al.}'s expression,~\cite{Egdell} the large $U$ and the large $g$
is related with each other though $\varepsilon_\infty$. This means that
the strong electron-phonon interaction is a consequence of strong
electron-electron interaction.

\section{Conclusion}
 
We have studied the electronic structure of VO$_2$ by measuring the
photoemission spectra of VO$_2$/TiO$_2$(001) thin films. The
metal-insulator transition was clearly observed as a change in the
spectral intensity at $E_{\rm F}$ accompanied by spectral weight
transfer between the coherent and incoherent parts of the spectral
function. From comparison of the ``bulk'' spectrum in the metallic
phase and the band-structure calculation, we have concluded that, while
the mass enhancement factor $m^*/m_b$ is almost unity, the momentum
dependence of the self-energy is important in the metallic phase,
possibly due to ferromagnetic fluctuations arising from the
multi-orbital character of the V $3d$ band. In the insulating phase, the
V $3d$ band is strongly broadened as the temperature is increased. We
attribute this temperature dependence to the electron-phonon interaction
and reproduced it using the independent boson model with a very large
coupling constant. This indicates that, while electron-electron
interaction is necessary to produce the band gap in the insulating
phase, electron-phonon interaction is important to fully understand the
electronic structure and charge transport in VO$_2$.

\section*{Acknowledgements}
The authors would like to thank A. Liebsch, S. Biermann and T. Mizokawa
for enlightening discussions. This work was supported by a Grant-in-Aid
for Scientific Research in Priority Area ``Novel Quantum Phenomena in
Transition Metal Oxides'' from the Ministry of Education, Culture,
Sports, Science and Technology, Japan.

\end{document}